\documentclass[preprint]{aastex}

\newcommand{\gtsima}{$\; \buildrel > \over \sim \;$}
\newcommand{\ltsima}{$\; \buildrel < \over \sim \;$}
\newcommand{\simgt}{\lower.5ex\hbox{\gtsima}}
\newcommand{\simlt}{\lower.5ex\hbox{\ltsima}}
\newcommand{\himpc}{{\hbox {$h^{-1}$}{\rm Mpc}} }
\newcommand{\bfx}{{\mbox{\boldmath $x$}}}
\newcommand{\bfth}{{\mbox{\boldmath $\theta$}}}

\shortauthors{Matsubara} 
\shorttitle{Gravitational Lensing and Correlation Function}

\begin{document}



\title{ THE GRAVITATIONAL LENSING IN REDSHIFT-SPACE CORRELATION
FUNCTIONS OF GALAXIES AND QUASARS }

\author{Takahiko Matsubara\altaffilmark{1,2}}
\affil{Department of Physics and Astronomy, 
        The Johns Hopkins University,
        3400 N.Charles Street, Baltimore, MD 21218
}
\email{matsu@pha.jhu.edu}

\altaffiltext{1}{Department of Physics, The University of Tokyo, Hongo
7-3-1, Tokyo 113-0033, Japan}
\altaffiltext{2}{Research Center for the Early Universe, Faculty of
        Science, The University of Tokyo, Tokyo 113-0033, Japan}

\begin{abstract}

The gravitational lensing, as well as the velocity field and the
cosmological light-cone warp, changes the observed correlation
function of high-redshift objects. We present an analytical expression
of 3D correlation function, simultaneously including those three
effects. When two objects are separated over several hundreds Mpc
along a line of sight, the observed correlation function is dominated
by the effect of gravitational lensing rather than the intrinsic
clustering. For a canonical lambda-CDM model, the lensing signals in
the galaxy-galaxy and galaxy-QSO correlations are beyond noise levels
in large-scale redshift surveys like the Sloan Digital Sky Survey.

\end{abstract}



\setcounter{equation}{0}
\section{INTRODUCTION}
\label{sec1}

The correlation function is one of the fundamental quantities in
searching into the physical origin of the universe. In large-scale
redshift surveys, redshifts and spherical positions on the sky of
luminous objects are used for estimating the spatial distribution of
mass, but the former is distorted by inhomogeneity of the universe.

Two intrinsic distortion effect on the correlation function originate
in the velocity field \citep{kai87} and the cosmological warp
\citep{bal96,mat96,mat00}. The former comes from the fact that an
observed redshift corresponds to the recession velocity which is
composed not only of the expansion of the universe but also of
peculiar velocities. The latter distortion is brought about by the
nonlinear mapping of the objects from the expanding curved space on a
light cone onto a flat redshift space \citep{alc79}.

In addition to those intrinsic distortions, there are secondary
distortions which are due to perturbations of light rays. While the
redshift is altered by the Sachs-Wolfe effect \citep{sac67}, the
spherical position is recast by the gravitational lensing
\citep{sch92,mel99,bar99}. Although the Sachs-Wolfe effect is not so
important, the gravitational lensing can affect the observable
correlation function in forthcoming redshift surveys as this effect is
efficient for high-redshift objects \citep{gun67}.

The effect of the gravitational lensing on the angular functions
$w(\theta)$ have been intensively investigated so far
\citep{bar99,moe98,kai92,vil96}. Recently, among others, cross
correlations of galaxies at different redshifts \citep{moe98} are
successfully applied to the commissioning data from the Sloan Digital
Sky Survey \citep{jai00}, to single out the weak lensing effect.

As for the 3D correlation function in redshift space, a qualitative
treatment of the lensing effect is given in \citet{sut99} to estimate
the upper limit of the effect, using the phenomenological Dyer-Roeder
distance \citep{dye73}. It is still not clear whether or not the
lensing is actually efficient where the intrinsic correlation function
is negligible on scales comparable to 100Mpc. The main purpose of this
{\em letter} is to give a quantitative treatment of this issue,
consistently including velocity and cosmological distortions, and
consequently to show the weak lensing actually has detectable effects
on 3D correlation function in redshift space when large-scale redshift
surveys like Sloan Digital Sky Survey (SDSS) are available.

\section{OBSERVABLE QUANTITIES}
\label{sec2}

We take the homogeneous, isotropic FRW metric with scalar
perturbations in longitudinal gauge:
\begin{eqnarray}
   ds^2 = 
   a^2(\tau)
   \left\{
      - (1 + 2\phi) d\tau^2
      + (1 - 2\phi)
      \left[
         d\chi^2
         + {S_K}^2(\chi)
         (d\theta^2 + \sin^2\theta d\varphi^2)
      \right]
   \right\},
\label{eq2-1}
\end{eqnarray}
where $a$ is the scale factor, and $\tau$ is the conformal time,
$d\tau = dt/a$, and $S_K$ is the comoving angular distance of the
spatial curvature $K = \Omega_0 + \lambda_0 - 1$. For example,
$S_K(\chi) = (-K)^{-1/2}\sinh\left[(-K)^{1/2}\chi\right]$ for open
universe, $K<0$, and $S_K(\chi) = \chi$ for flat universe, $K=0$. We
adopt the unit $c = H_0 = 1$ throughout this letter.

{}From the first-order Einstein equation of the metric in equation
(\ref{eq2-1}), the density contrast $\delta(\bfx,\tau)$ and the
velocity field $v^i(\bfx,\tau)$ on scales much less than the curvature
scale satisfy
\begin{eqnarray}
   \triangle\phi =
   \frac{3\Omega_0}{2} \frac{\delta}{a},
\qquad
   v^i = \frac{2}{3 \Omega_0} a H f
	 \nabla^i \phi,
\label{eq2-3}
\end{eqnarray}
where $H = \dot{a}/a$, $f = d\ln D/d\ln a$, and $D$ is the linear
growth rate \citep{pee80}. The Laplacian $\triangle$ is taken with
respect to comoving coordinates.

Let us consider a light ray emitted from an object at comoving
coordinates $(\chi,\theta,\varphi;\tau)$, which an observer recieves
at $\tau_0$. The conventional redshift $z = a^{-1} - 1$ is given by
$\chi = \int_0^z dz H^{-1}$, but the actually observed redshift is
changed by the line-of-sight peculiar velocity, $V = n_i v^i$, where
$n_i$ is a line-of-sight unit vector, and also by the gravitational
potential, $\phi$. {}From the time-component of the geodesic equation
of the light ray, the observed redshift $z_{\rm s}$ is given by
\citep{sac67}
\begin{eqnarray}
   z_{\rm s} = z + (1 + z) \left[ V(\chi) - V(0) - \phi(\chi) +
   \phi(0) - 2 \int_{\tau_0-\chi}^{\tau_0} d\tau
   \frac{\partial\phi}{\partial\tau} \right],
\label{eq2-6}
\end{eqnarray}
where we abbreviate the function on a light cone as $V(\chi) \equiv
V(\chi,\theta,\varphi;\tau_0-\chi)$, and so as $\phi(\chi)$. The
integral is performed on the light cone for a fixed direction of line
of sight.

Now, we consider the small angle approximation so that light rays are
confined to a narrow cone around the polar axis, $\theta \ll 1$, and
introduce new coordinates $\theta_1 = \theta\cos\varphi$, $\theta_2 =
\theta\sin\varphi$, following \citet{kai98}. Adopting the Born
approximation, the angular components of the geodesic equation reduces
to equation for the observed angular components $\theta_{{\rm s}a}$
($a = 1,2$):
\begin{eqnarray}
   \theta_{{\rm s}a} =
   \theta_a +
   \frac{2}{S_K(\chi)}
   \int_0^{\chi} d\chi' S_K(\chi - \chi')
   \partial_a \phi(\chi'),
\label{eq2-7}
\end{eqnarray}
where $\partial_a = {S_K}^{-1}\partial/\partial\theta_a$.

The apparent luminosity of the light is magnified by a factor $A(z) =
|\det (\partial\theta_{{\rm s}b}/\partial\theta_a)| = 1 + 2\kappa$,
where $\kappa$ is a local convergence field of weak lensing
\citep{sch92,ber97,kai98} for a fixed redshift of source object:
\begin{eqnarray}
   \kappa(z,\bfth) = 
   \int_0^\chi d\chi' g(\chi,\chi')
   \partial_a \partial_a \phi(\chi'); \quad
   g(\chi,\chi') \equiv 
  \frac{S_K(\chi') S_K(\chi-\chi')}{S_K(\chi)}.
\label{eq2-11}
\end{eqnarray}
Due to the magnification, the observed apparent magnitude $m_{\rm s}$
is given by $m_{\rm s} = m - 2.5\log_{10}A = m - 5\kappa/\ln10$
\citep{bro95}, where $m$ is the apparent magnitude in the absence of
lensing.

The magnitude-limited number density in real space $n_{\rm
r}(z,\bfth;<\!\!m)$ and that in observed redshift space $n_{\rm
s}(z_{\rm s},\bfth_{\rm s};<\!\!m_{\rm s})$ are related by the number
conservation equation, $n_{\rm r} z^2 dz d^2\theta = n_{\rm s} {z_{\rm
s}}^2 dz_{\rm s} d^2\theta_{\rm s}$, while the number density is given
by $n(z,\bfth;<\!\!m) = [1 + \delta(z,\bfth)]N(z;<\!\!m)/(4\pi z^2)$,
where $\delta(z,\bfth)$ is the density contrast and $N(z;<\!\!m)$ is
the magnitude-limited number count per redshift. Evaluating the
Jacobian, one obtains the relation between $n$ and $n_{\rm s}$, as
well as the relation between density contrasts. The result contains
the terms with $\delta$, $V$, $\phi$ and their derivatives. For
fluctuations on a scale $k$ in units of Hubble distance, such
variables scales as $V \sim k^{-1}\delta$, $\partial V \sim \delta$,
$\phi \sim k^{-2}\delta$, $\partial\phi \sim k^{-1}\delta$, and
$\partial^2\phi \sim \delta$, where $\partial$ schematically
represents the spatial derivatives in comoving coordinates.
Consistently to the small angle approximation, we neglect the
fluctuations which scale as $k^{-1}$ and $k^{-2}$, because $k$ is
large enough on scales below Hubble distance. Eventually, the
distorted density contrast is given by
\begin{eqnarray}
   \delta_{\rm s} =
   \delta_{\rm r} -
   \frac{1+z}{H}\frac{\partial V}{\partial\chi} +
   (5\alpha - 2)\,\kappa
\label{eq2-10}
\end{eqnarray}
where $\delta_{\rm r}(z,\bfth)$ is the number density contrast of the
objects in real space, and $H(z)$ is the Hubble parameter at $z$. The
logarithmic slope of the number counts $\alpha$ at the limiting
magnitude $m$ is given by $\alpha(z,m) =
\partial\log_{10}N(z;<\!\!m)/\partial m$ (c.f., Moessner, Jain \&
Villumsen 1998). The first two terms of equation (\ref{eq2-10})
depends strongly on radius $z$ and angles $\bfth$, while the last, the
lens surface density, depends strongly on angles but very weakly on
radius. The first term of equation (\ref{eq2-10}) is the real density
fluctuations, the second term is the velocity distortion
\citep{kai87,mat96}, and the last term is the effect of weak lensing,
which consists of the contribution from the modulation by
magnification bias, $5\alpha\kappa$, and of the alternation of surface
density by lensing, $-2\kappa$. In the following, we denote each term
in equation (\ref{eq2-10}) as $\delta_{\rm s} = \delta_{\rm r} +
\delta_{\rm v} + \delta_{\rm l}$.

\section{CORRELATION FUNCTION}
\label{sec3}

We consider the correlation function between two objects
$(z_1,\bfth_1)$ and $(z_2,\bfth_2)$ in the small angle approximation,
$\theta \equiv |\bfth_1 - \bfth_2| \ll 1$, and assume $z_1 \leq z_2$
without loss of generality. In the absence of the lensing term
$\delta_{\rm l}$, the correlation function is given by \citet{mat96},
which generalize the work by \citet{ham92} to high-redshift objects.
In the coordinate system $(z_1, z_2, \theta)$, their result, with
slight modification allowing difference of the kind of two objects, is
expressed as follows:
\begin{eqnarray} 
&&
   \xi_{\rm MS} =
   \left\langle
      \left[\delta_{\rm r}(1) + \delta_{\rm v}(1)\right]
      \left[\delta_{\rm r}(2) + \delta_{\rm v}(2)\right]
   \right\rangle
\nonumber\\
&&\qquad =
   \left[
      1 + \frac13 (\beta_1 + \beta_2) + \frac15 \beta_1 \beta_2
   \right]
   \xi_0(x;\bar{z}) P_0(\mu)
   \nonumber\\
   && \qquad\qquad -\,
   \left[
      \frac23 (\beta_1 + \beta_2) + \frac47 \beta_1 \beta_2
   \right]
   \xi_2(x;\bar{z}) P_2(\mu)
   + \frac{8}{35} \beta_1 \beta_2 \xi_4(x;\bar{z}) P_4(\mu),
\label{eq3-1}
\end{eqnarray}
where $\bar{z} = (z_1 + z_2)/2$, $\beta_i = f(z_i)/b_i(z_i) \simeq
\Omega^{0.6}(z_i)/b_i(z_i)$, and $b_i(z_i)$ is the bias parameter of
object $i$ ($i = 1,2$) at redshift $z_i$. Similarly, a bar for any
variable means the evaluation at $\bar{z}$ and subscripts $1,2$ assume
evaluation at objects 1 and 2, respectively. We denote the comoving
separation $x \equiv \sqrt{[S_K(\bar{\chi})]^2 \theta^2 + (\chi_2 -
\chi_1)^2}$, the comoving cosine $\mu\equiv (\chi_2 - \chi_1)/x$, and
$P_n$'s are Legendre polynomials, and
\begin{eqnarray} 
   \xi_{2l}(x;\bar{z}) =
   \bar{b}^2
   \int_0^\infty \frac{k^2dk}{2\pi^2} j_{2l}(kx) P(k;\bar{z}),
\label{eq3-2}
\end{eqnarray}
where $P(k;z)$ is the power spectrum at redshift $z$. This formula is
valid only for distant observer approximation, $x \ll \chi_1$ and
$\theta \ll 1$. We have omitted the finger-of-God effect which is only
important on scales less than $10\himpc$ or $1000{\rm km/s}$. In the
following, we are interested in the scales of 30Mpc or larger where
lensing effect appears, so that we can safely ignore the nonlinear
effect like finger-of-God effect.

Adopting the small angle approximation (e.g., Bernardeau et al. 1997;
Kaiser 1998; Moessner et al. 1998), the correlations involving the
lensing term $\delta_{\rm l}$ are obtained as
\begin{eqnarray}
&&
   \xi_{\rm rl} = 
   \left\langle
      \delta_{\rm r}(1) \delta_{\rm l}(2)
   \right\rangle =
   \frac32 \Omega_0
   b_1 (5\alpha_2 - 2)
   g(\chi_2,\chi_1)
   (1 + z_1)
   \xi_{\rm p}\left[\theta S_K(\chi_1);z_1\right],
\label{eq3-4a}\\
&&
   \xi_{\rm ll} =
   \left\langle
      \delta_{\rm l}(1) \delta_{\rm l}(2)
   \right\rangle =
   \left(\frac32 \Omega_0\right)^2
   (5\alpha_1 - 2) (5\alpha_2 - 2)
\nonumber\\
&&\qquad\qquad\qquad\qquad\qquad\,\times
   \int_0^{\chi_1} d\chi(z)
   g(\chi_1,\chi) g(\chi_2,\chi)
   (1 + z)^2
   \xi_{\rm p}\left[\theta S_K(\chi);z\right],
\label{eq3-4c}
\end{eqnarray}
where $d\chi(z) = dz/H(z)$, $\xi_{\rm p}$ is the projected correlation
function defined by
\begin{eqnarray}
   \xi_{\rm p}(y;z) =
   \int_{-\infty}^\infty dx
   \xi\left(\sqrt{x^2 + y^2};z \right) =
   \int_0^\infty \frac{kdk}{2\pi} J_0(ky)P(k;z),
\label{eq2-16}
\end{eqnarray}
and $\xi(r;z) = \xi_0(r;z)$ is the correlation function in real space
at $z$. The terms $\langle\delta_{\rm l}(1) \delta_{\rm r}(2)\rangle$,
$\langle\delta_{\rm l}(1) \delta_{\rm r}(2)\rangle$ are zero for $z_1
\leq z_2$ in the small angle approximation. The velocity-lensing term,
$\xi_{\rm vl} =\langle \delta_{\rm v}(1) \delta_{\rm l}(2)\rangle$ is
explicitly calculated to be zero, which is because the term
$\delta_{\rm v}$ only depends on fluctuations along the line of sight
which are smoothed out. If we convolve the above expression of
$\xi_{\rm rl}$ and $\xi_{\rm ll}$ with a selection function along the
line of sight, we obtain the form of angular correlation function with
the effect of weak lensing \citep{bar95,vil96,dol97,moe98,moe98a}.

Thus the total correlation function in 3D redshift space is given by
$\xi_{\rm tot}(z_1,z_2,\theta) = \xi_{\rm MS} + \xi_{\rm rl} +
\xi_{\rm ll}$ for $z_1 \leq z_2$ and $\theta \ll 1$. The first term
dominates on scales much smaller than Hubble distance, while the last
two terms dominate on scales comparable to the Hubble distance along
the line of sight. Therefore, even though $\xi_{\rm MS}$ is valid only
for $z_1-z_2 \ll z_1$ (distant observer approximation), we can use the
form $\xi_{\rm tot}$ even when $z_1-z_2 \sim z_1$.

In Figure~\ref{fig1}, we plot the total correlation function $\xi_{\rm
tot}$ together with each component, $\xi_{\rm MS}$, $\xi_{\rm rl}$,
and $\xi_{\rm ll}$. With the choice of the CDM-like initial power
spectrum \citep{bar86} with a shape parameter $\Gamma = 0.25$ and a
linear amplitude $\sigma_8 = 1$, we use the fitting formula for the
fully non-linear power spectrum of \citet{pea96} for lensing
correlations, $\xi_{\rm rl}$ and $\xi_{\rm ll}$. Linear predictions
are also plotted in the lower panels. The nonlinearity of the
intrinsic correlation $\xi_{\rm MS}$, which is only important for the
region $z_2-z_1 \simlt 0.003$, is ignored. We exemplify the
low-density flat model with $\Omega_0 =0.3$, $\lambda = 0.7$. In the
upper panels, the slope of the number counts is assumed as $\alpha =
1$, and the bias factor as $b = 1$, regardless of the redshift. This
example corresponds to $z \sim 0.2$ and $m \sim 18$ of galaxies as
seen in Table~\ref{tab1}, in which the slope $\alpha$ is calculated
from the B-band luminosity function of APM galaxies \citep{lov92}, and
of quasars \citep{boy88}. The limiting magnitudes assumed in
Table~\ref{tab1} correspond to estimated SDSS redshift data of
galaxies and quasars. In practice, the slope $\alpha$ can be
observationally determined for individual catalogue of quasar or
galaxy redshift surveys. Each component of correlations roughly scales
as $\xi_{\rm MS} \propto {\sigma_8}^2 b_1 b_2$, $\xi_{\rm rl} \propto
{\sigma_8}^2 \Omega_0 b_1 (5\alpha_2-2)$, $\xi_{\rm ll} \propto
{\sigma_8}^2 {\Omega_0}^2 (5\alpha_1-2)(5\alpha_2-2)$ for other
parameters and models. \placetable{tab1}

In the lower panels of Figure~\ref{fig1}, the galaxy-galaxy (G-G),
galaxy-QSO (G-Q), and QSO-QSO (Q-Q) correlations are plotted,
assuming the SDSS slope $\alpha$ of Table~\ref{tab1}. The bias
parameter is set $b=1$ and $3$ for galaxies and quasars, respectively.
The line-of-sight separations are large enough in lower panels, so
that the intrinsic clustering is negligible.

\section{DISCUSSION}
\label{sec4}

The absolute value of the intrinsic clustering component $\xi_{\rm
MS}$ is a decreasing function of the line-of-sight separation, $z_2 -
z_1$, except the vicinity of zero crossings. On the other hand, the
lens-lens component $\xi_{\rm ll}$ is almost independent on the
separation and the density-lens component $\xi_{\rm rl}$ is an
increasing function. Those behaviors are understood by the fact that
the weak lensing is efficient between the object and the observer.
Thus, intrinsic clustering component dominates for small separations,
while lens-lens and/or density-lens components dominates for large
separations along the direction of line of sight.

\begin{figure}
\epsscale{0.88} \plotone{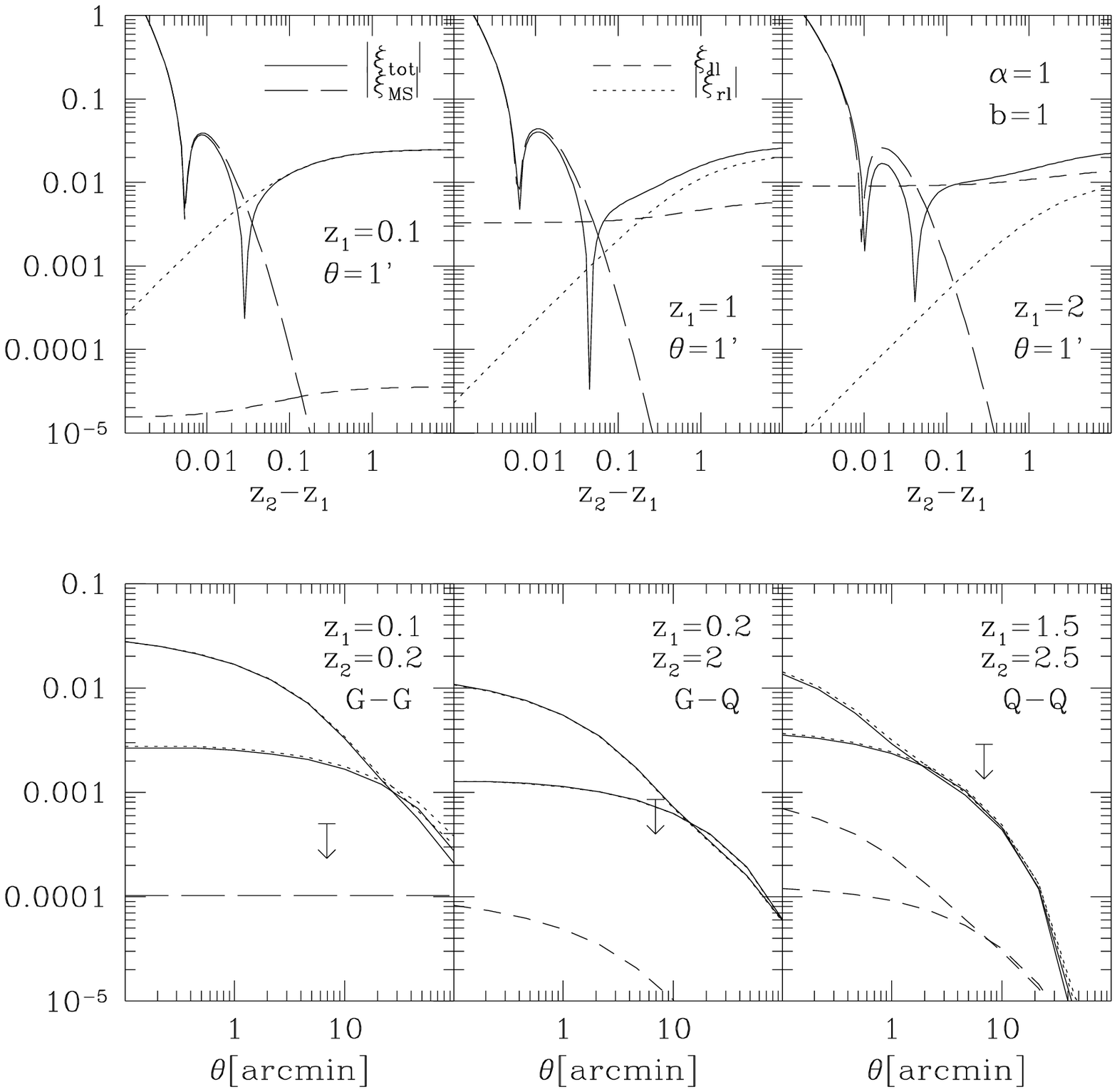} \figcaption[fig1.eps]{The
correlation function along the line of sight for a flat LCDM model.
Long-dashed lines: intrinsic clustering, $\xi_{\rm MS}$, dashed lines:
lens-lens correlation, $\xi_{\rm ll}$, dotted lines: density-lens
correlation, $\xi_{\rm rl}$, solid lines: total correlation function,
$\xi_{\rm tot}$. In the upper panels, $z_1$ and $\theta$ are fixed in
each panel, with $\alpha=1$, $b=1$. In lower panels, $z_1$ and $z_2$
are fixed and parameters which mimic the SDSS redshift catalogue (see
text) are assumed. The noise levels for the SDSS are also shown. From
left to right are plotted galaxy-galaxy, galaxy-QSO, and QSO-QSO
correlations. Nonlinear predictions are plotted except the lines which
are not enhanced on small angles in lower panels.
\label{fig1}}
\end{figure}

The lower panels in Figure~\ref{fig1} show the region where lensing
contribution dominates in the case of the SDSS magnitude limits to
illustrate the typical magnitude of correlations. Are those lensing
signals detectable? The statistical uncertainty in estimating the
correlation function is given by $(\delta\xi)^2 = \Omega/(\delta\Omega
N_1 N_2)$ \citep{pee80}, where $N_1$ and $N_2$ are numbers of object
in the bin used for redshifts $z_1$ and $z_2$, respectively, $\Omega$
is the solid angle subtended by the survey area, and $\delta\Omega$ is
the fraction in the bin used for angle $\theta$. To increase the
signal to noise ratio, it is desirable to use large bins for $\theta$.
To be specific, we consider a bin $[1',10']$, and theoretical curves
are integrated accordingly, so that $\delta\Omega \sim 100 \pi\
[\mbox{arcmin}^2]$. The effective scale of this bin is given by
$\theta_{\rm eff} = 10'/\sqrt{2}$ for $\xi \propto \theta^{-1}$. In
the SDSS, $\Omega \sim \pi\ \mbox{[str]} \sim 1.2 \pi \times 10^7\
[\mbox{arcmin}^2]$, and the estimated numbers of galaxies and quasars
are $N_{\rm G} = 10^6$ and $N_{\rm Q} = 1.7 \times 10^5$,
respectively. Assuming we take sufficiently large bins of redshifts
(this choice is similar to considering angular correlation functions),
the consequent estimates of the statistical error are given by $5.0
\times 10^{-4}$ for G-G, $8.5\times 10^{-4}$ for G-Q, and $2.9 \times
10^{-3}$ for Q-Q correlations, which are plotted in lower panels. The
S/N ratios turn out to be about $10$, $1.3$, and $0.22$ for G-G, G-Q,
and Q-Q correlations, respectively.

Therefore, the weak lensing in 3D correlation function of galaxies in
the SDSS is definitely detectable, and the detection of galaxy-QSO
cross-correlation is marginal, while the quasar correlation by lensing
is below the noise level in the SDSS. In order to detect the QSO-QSO
lensing effect, the sample should be at least 5 times larger than the
SDSS, or parameters $\sigma_8$, $\Omega_0$, $b$, $\alpha$ should be
larger than assumed values.

In summary, we have obtained a theoretical prediction of correlation
function in redshift space, taking into account the effect of weak
lensing, together with velocity distortions and cosmological
distortions on a light-cone. Each effect contributes differently to
the correlation function, and is realistically detectable. Our result
provides a fundamental link between theoretical models and the
observed correlation function in the 3D redshift survey data. Besides
the determination of the power spectrum itself, various cosmological
parameters, especially the bias parameter, can be estimated by proper
likelihood analyses, including KL transform of the correlation matrix
\citep{vog96,mat00a}. One may also be tempted to assume the
cosmological parameters before analysing data. In which case, the
error originated in choosing the wrong cosmological model is roughly
given by the order of the redshift $z$ times the error of cosmological
parameters, since the Alcock-Paczy\'nski effect is roughly
proportional to $z$ up to $z=1$-$2$.

\acknowledgements

I am grateful to Bhuvnesh Jain for many helpful discussions. I would
like to thank Yasushi Suto and Alex Szalay for stimulating
discussions. I acknowledge support from JSPS Postdoctoral Fellowships
for Research Abroad.

\clearpage


\bigskip

\clearpage

\begin{deluxetable}{ccccccc|cccccc}
\tablecolumns{13}
\tablewidth{0pc}
\tablecaption{Slope $\alpha(z,m)$ of the number counts with fixed
redshifts. \label{tab1}} 
\tablehead{
\colhead{} & \multicolumn{6}{c|}{Galaxies\tablenotemark{a} ($B_{\rm lim} =
18.8$)} & \multicolumn{6}{c}{Quasars\tablenotemark{b} ($B_{\rm lim} =
20.0$)}}
\startdata
$z$... & 0.05 & 0.10 & 0.15 & 0.20 & 0.25 & 0.30 & 0.5 & 1.0 & 1.5 &
 2.0 & 2.5 & 3.0 \\
$\alpha$... & 0.20 & 0.42 & 0.75 & 1.2 & 1.8 & 2.6 & 0.26 & 0.29 &
 0.30 & 0.29 & 0.29 & 0.28 \\
\enddata
\tablenotetext{a}{B-band luminosity function of the APM galaxies
\citep{lov92} is assumed}
\tablenotetext{b}{B-band luminosity function of the quasar sample
\citep{boy88} is assumed}
\end{deluxetable}

\end{document}